\documentstyle[10pt,emulateapj]{article}

\renewcommand{\tabcolsep}{3pt}

\newcommand{\NH}{{$N_{\rm H}$}}
\newcommand{\LX}{{$L_{\rm X}$}}
\newcommand{\LB}{{$L_B$}}
\newcommand{\LFIR}{{$L_{\rm FIR}$}}
\newcommand{\LHa}{$L_{{\rm H} \alpha}$}

\newcommand{\hii}{\ion{H}{2}}
\def\deg{$^{\circ}$}
\newcommand{\amin}{{$^{\prime}$}}
\newcommand{\asec}{{$^{\prime\prime}$}}
\newcommand{\eps}{ergs s$^{-1}$}
\newcommand{\pcm}{cm$^{-2}$}

\newcommand{\solmass}{$M_\odot$}

\newcommand{\asca}{{\it ASCA}}
\newcommand{\Einstein}{{\it Einstein}}
\newcommand{\rosat}{{\it ROSAT}}
\newcommand{\HST}{{\it HST}}

\newcommand{\gtsima}{$\; \buildrel > \over \sim \;$}
\newcommand{\simgt}{\lower.5ex\hbox{\gtsima}}
\newcommand{\ltsima}{$\; \buildrel < \over \sim \;$}
\newcommand{\simlt}{\lower.5ex\hbox{\ltsima}}

\slugcomment{Accepted for publication in {\it The Astrophysical Journal}.}

\lefthead{Terashima et al.}
\righthead{\asca\ Observations of ``Type 2'' LINERs}

\begin{document}

\title{{\it ASCA} Observations of ``Type 2'' LINERs: Evidence for a Stellar 
Source of Ionization}

\author{Yuichi Terashima}
\affil{NASA Goddard Space Flight Center, Code 662, Greenbelt, MD 20771
}

\author{Luis C. Ho}
\affil{Carnegie Observatories, 813 Santa Barbara St. Pasadena, CA 91101-1292}

\author{Andrew F. Ptak}
\affil{Department of Physics, Carnegie Mellon University, 5000 Forbes Ave., Pittsburgh, PA 15213}

\author{Richard F. Mushotzky, Peter J. Serlemitsos, and Tahir Yaqoob
\altaffilmark{1}}
\affil{NASA Goddard Space Flight Center, Code 662, Greenbelt, MD 20771
}

\altaffiltext{1}{Universities Space Research Association.}

\and

\author{Hideyo Kunieda\altaffilmark{2}}
\affil{Department of Physics, Nagoya University, Chikusa-ku, Nagoya, 
464-8602, Japan}

\altaffiltext{2}{present address: Institute of Space and Astronautical
Science, Yoshinodai 3-1-1, Sagamihara, Kanagawa 229-8510, Japan}

\begin{abstract}

We present \asca\ observations of LINERs without broad H$\alpha$
emission in their optical spectra.  The sample of ``type 2'' LINERs
consists of NGC 404, 4111, 4192, 4457, and 4569. We have detected X-ray
emission from all the objects except for NGC 404; among the detected
objects are two so-called transition objects (NGC 4192 and NGC 4569),
which have been postulated to be composite nuclei having both an \hii\
region and a LINER component.

The images of NGC 4111 and NGC 4569 in the soft (0.5--2 keV) and hard
(2--7 keV) X-ray bands are extended on scales of several kpc.  The
X-ray spectra of NGC 4111, NGC 4457 and NGC 4569 are well fitted by a
two-component model that consists of soft thermal emission with
$kT\sim0.65$ keV and a hard component represented by a power law
(photon index $\sim$ 2) or by thermal bremsstrahlung emission ($kT$
$\sim$ several keV). The extended hard X-rays probably come from
discrete sources, while the soft emission most likely originates from
hot gas produced by active star formation in the host galaxy.  We have
found no clear evidence for the presence of active galactic nuclei
(AGNs) in the sample. Using black hole masses estimated from host
galaxy bulge luminosities, we obtain an upper limit on the implied
Eddington ratios less than $5\times10^{-5}$. If an AGN component is
the primary ionization source of the optical emission lines, then it
must be heavily obscured with a column density significantly larger
than $10^{23}$ {\pcm}, since the observed X-ray luminosity is
insufficient to drive the luminosities of the optical emission
lines. Alternatively, the optical emission could be ionized by a
population of exceptionally hot stars.  This interpretation is
consistent with the small [O~I] $\lambda$6300/H$\alpha$ ratios
observed in these sources, the ultraviolet spectral characteristics in
the cases where such information exists, and the X-ray results
reported here.

We also analyze the X-ray properties of NGC 4117, a low-luminosity
Seyfert 2 galaxy serendipitously observed in the field of NGC 4111.

\end{abstract}


\section{Introduction}

LINERs (low-ionization nuclear emission-line regions; Heckman 1980)
are found in a significant fraction of bright galaxies (Ho,
Filippenko, \& Sargent 1997a). The ionizing source of LINERs is still
under debate (see Filippenko 1996 for a review), with candidate
ionization mechanisms being photoionization by low-luminosity active
galactic nuclei (AGNs), photoionization by very hot stars, and
collisional ionization by fast shocks.  Recent observations have shown
that at least some LINERs are low-luminosity AGNs (hereafter LLAGNs);
see Ho (1999a and references therein) for a review. According to the
extensive spectroscopic survey of Ho et al. (1997b), which includes
all bright ($B_T$ {\simlt} 12.5 mag) galaxies with declinations
greater than 0\deg, about 20\% of LINERs exhibit a broad H$\alpha$
emission line in their optical spectra.  Hard X-ray observations
provide a powerful means for searching for evidence of an AGN by
detecting a pointlike hard X-ray source at the nucleus.  Compact hard
X-ray sources have been detected in LINERs that show broad H$\alpha$
emission (hereafter LINER 1s; Iyomoto et al. 1996, 1998; Ptak et
al. 1998; Terashima et al. 1998a; Guainazzi \& Antonelli 1999; Weaver
et al. 1999). The X-ray spectra of these objects are well represented
by a two-component model: a power-law component plus soft thermal
emission. The X-ray spectra of the hard component are quite similar to
those of Seyfert galaxies, and typical X-ray luminosities are
$10^{40-41}$~{\eps}. The H$\alpha$ luminosities of LINER 1s are
positively correlated with the X-ray luminosities in the 2--10 keV
band (Terashima 1999; Terashima et al. 1999). These observations
strongly support the idea that most LINER 1s are LLAGNs.

LINERs constitute the majority of the objects that show spectroscopic
evidence for nuclear activity, and most LINERs ($\sim80$\%; Ho et
al. 1997a, b) show no detectable broad H$\alpha$ emission and are
classified as ``LINER 2s''. Therefore, LINER 2s are the most abundant
form of low-level activity in nearby galaxies. It is not clear whether
the origin of LINER 2s is similar to that of LINER 1s. If LINER 2s are
also genuine AGNs, then the emission from their nuclei may be
obscured, by analogy with the popular obscuration model for Seyfert 2
galaxies (Lawrence \& Elvis 1982; Antonucci \& Miller 1985).  If this
is the case, then the X-ray spectra of LINER 2s should show evidence
for heavy absorption ({\NH} $>10^{23}$~{\pcm}) and strong fluorescent
iron K emission lines.  For example, NGC 1052, a LINER 1.9 (Ho et
al. 1997a) from which polarized broad H$\alpha$ line has been detected
(Barth et al. 1999), shows an X-ray spectrum absorbed by a column
density of {\NH} $\approx\,3\times10^{23}$~{\pcm} and a fluorescent
iron~K emission line with an equivalent width of $\sim$300 eV (Weaver
et al. 1999). These X-ray characteristics are quite similar to those
of luminous Seyfert 2 galaxies (Awaki et al. 1991a; Smith \& Done
1994; Turner et al. 1997a). Thus, this is an example of an active
nucleus which is a low-ionization analog of Seyfert 2 galaxies such as
NGC 1068.

Alternatively, the optical emission lines in LINER 2s may be ionized
by sources other than an AGN. Collisional ionization from fast-moving
shocks (e.g., Koski \& Osterbrock 1976; Fosbury et al. 1978; Dopita \&
Sutherland 1995) and photoionization by a cluster of hot, young stars
(Terlevich \& Melnick 1985; Filippenko \& Terlevich 1992; Shields
1992) have also been proposed as possible excitation mechanisms to
power the narrow-line emission in LINERs.  Ultraviolet (UV) spectra of
several LINER 2s are available from the {\it Hubble Space Telescope}
({\HST}), and these indicate the presence of massive stars in some
LINER 2s (Maoz et al. 1998). Maoz et al. (1998) find that in NGC 404,
NGC 4569, and NGC 5055 hot stars play a significant role as an
ionizing source for the optical emission lines.  In order to explain
the observed H$\alpha$ luminosities by stellar photoionization,
however, very massive stars ($M\,\approx\, 100$ \solmass) are required
to be still present (see Fig. 5 in Maoz et al. 1998).

Although UV spectroscopy can probe the presence of massive stars, only
UV bright objects ($\sim$20\%--30\% of LINERs; Maoz et al. 1995; Barth
et al. 1998; Ho, Filippenko, \& Sargent 1999) can be studied.
Moreover, Maoz et al. (1998) have argued that even in objects such as
M81, NGC 4579 or NGC 4594, where an LLAGN is known to be present based
on other evidence, the observed UV continuum power is insufficient to
account for the observed line emission, and the nonstellar component
most likely becomes more prominent at higher energies. Searching for
the ionizing source in the X-rays is necessary to test this
hypothesis.

Only a limited number of X-ray observations of LINER 2s have been
performed so far. Previous X-ray observations with {\Einstein} and
{\rosat} were limited to soft energies, where heavily obscured AGNs
are difficult to detect. Furthermore, the limited spectral resolution
and bandpass of these observations cannot distinguish the thermal
emission of the host galaxy from the emission from the AGN.

We observed a small sample of three LINER 2 nuclei (NGC 404, NGC 4111,
and NGC 4457) with the {\asca} satellite to search for a hidden
ionizing source; the sample also included NGC 4192 and NGC 4569, which
are classified as ``transition objects,'' emission-line nuclei whose
optical spectra suggest a composite source of ionization, possibly due
to a genuine LINER nucleus mixed in with signal from circumnuclear
\hii\ regions.  The imaging capability of {\asca} (Tanaka, Inoue, \&
Holt 1994), which extends up to 10 keV, and its moderate spectral
resolution enable it to identify thermal emission from the host
galaxy.  We also analyze the X-ray properties of NGC 4117, a
low-luminosity Seyfert 2 galaxy serendipitously observed in the field
of NGC 4111.

This paper is organized as follows. In $\S2$ we summarize the {\it
ASCA} observations and data reduction. Image and spectral analysis are
reported in $\S3$ and $\S4$, respectively. We discuss the origin of
X-ray emission and the ionizing source in type 2 LINERs in $\S5$. A
summary of our findings is presented in $\S6$.

\section{Observations}

We observed the three LINER 2s and two transition objects shown in
Table 1. These objects are selected from the Palomar survey of nearby
galaxies conducted by Ho et al. (1995, 1997a), for which Ho et
al. (1997b) determined that broad H$\alpha$ emission is not
present. We selected objects that are bright in the narrow H$\alpha$
emission line, since a large X-ray flux is expected if the ionization
source is due to an AGN (Halpern, \& Steiner 1983; Elvis, Soltan, \&
Keel 1984; Ward et al. 1988; Koratkar et al. 1995; Terashima 1999;
Terashima et al. 1999).  We also gave preference to objects that have
previously been studied in the UV using the {\HST}. Both NGC 404 and
NGC 4569 are UV bright and have been studied spectroscopically in the
UV by Maoz et al. (1998), while NGC 4111 and NGC 4192 were imaged in
the UV but were not detected (Maoz et al. 1996; Barth et al. 1998).

\placetable{tbl-1}

The log of the {\asca} observations is shown in Table 2. Detailed
descriptions of the {\asca} instruments can be found in Serlemitsos et
al. (1995), Ohashi et al. (1996), Makishima et al. (1996), Burke et
al. (1994), and Yamashita et al. (1997). The observation mode of the
Solid-state Imaging Spectrometers (SIS) is summarized in Table 2; the
Gas Imaging Spectrometers (GIS) were operated in the nominal
pulse-height mode. We screened the data using standard criteria.  We
excluded data taken when (1) the elevation angle from the earth's limb
was less than 5\deg\ , (2) the cut-off rigidity was less than 6 GeV
c$^{-1}$, (3) the satellite was passing through the South Atlantic
Anomaly, and (4) the elevation angle from the day earth's limb was
less than 25\deg\ (only for the SIS). The observed count rates, after
background subtraction, and the net exposure times, after data
screening, are also tabulated in Table 2. Although we observed NGC
404, 4111, 4192 and 4569 on two occasions in order to search for
variability, no significant variability was found. The typical upper
limit on variability is 50\%. We therefore use images and spectra
combined from the two observations in the following analysis. In this
paper, the quoted errors are at the 90\% confidence level for one
parameter of interest, unless otherwise noted.

\placetable{tbl-2}

\section{X-ray images}

We detected X-ray emission from all objects except for NGC 404.  In
this section, we show X-ray images and estimate the spatial extension
of the X-ray emission.

\subsection{NGC 404}

The nucleus of NGC 404 was not detected in either the SIS or in the GIS
images. One serendipitous source was detected in the SIS image in the
0.5--2 keV band of the second observation. The SIS image in the 0.5--2
keV band is shown in Figure 1{\it a}. The peak position of this source is
($\alpha$,~$\delta$)$_{\rm J2000}$ = ($1^{h}\,9^{m}\,28^{s}$, 35\deg\ 38\amin\ 
59\asec), and the error radius is about 1$^{\prime}$.  This source is not 
clearly seen in the GIS image or in the SIS image above 2 keV.

We calculated an upper limit for the X-ray flux seen toward the
nucleus of NGC 404 using the following procedure. We made a
one-dimensional projection of width 2\farcm5 along the nucleus and the
serendipitous source and then fitted the profile with a model
consisting of two point-spread functions (PSFs) at the positions of
the two objects plus a constant background. The model PSFs were
obtained using a ray-tracing code, and they were projected using the
same method applied to the data. The free parameters are the
normalizations of the two PSFs and the background level. We fitted the
projected SIS images in the 0.5--2 keV and 2--10 keV bands and
obtained a 3$\sigma$ upper limit of $1.1\times10^{-14}$ ergs s$^{-1}$
cm$^{-2}$ and $6.6 \times10^{-14}$ ergs s$^{-1}$ cm$^{-2}$,
respectively, assuming a power-law spectrum with a photon index of
$\Gamma$ = 2.  For an assumed distance of 2.4 Mpc (Tully 1988), these
upper limits for the X-ray flux correspond to luminosities of
$7.6\times10^{36}$ ergs s$^{-1}$ in the 0.5--2 keV band and
$4.6\times10^{37}$ ergs s$^{-1}$ in the 2--10 keV band.

\subsection{NGC 4111}

Figures 1{\it c} and 1{\it d} show the GIS images in the 0.5--2 keV
and 2--7 keV bands, respectively. At least three X-ray sources were
detected in the GIS field of view, and the positions and tentative
identifications of these are summarized in Table 3. One of these
sources, at ($\alpha$,~$\delta$)$_{\rm J2000}$ =
($12^{h}\,7^{m}\,47^{s}$, 43\deg\ 6\amin\ 53\asec) is brighter in the
$>$2 keV image than in the $<$2 keV image.  This hard source is
positionally coincident, within the astrometric uncertainty of \asca,
with NGC 4117, a $B\,\approx$ 14.0 mag Seyfert 2 galaxy first
recognized by Huchra, Wyatt, \& Davis (1982).  The other two sources
are brighter in the soft-band image. One of the soft sources coincides
with the nucleus of NGC 4111, while the other has no counterpart in
the NASA Extragalactic Database (NED).

NGC 4111 is detected also in the SIS image, but the other sources are
out of the field of view of the detector. In order to estimate the
spatial extent of the X-ray emission, we made azimuthally averaged
radial profiles of surface brightness using the SIS images and
compared them with those of the PSF. We used the SIS images to examine
the spatial extent of the emission, since the spatial resolution of
the SIS is better than that of the GIS.  We tried to fit the radial
profiles in the 0.5--2 keV and 2--7 keV bands with a PSF plus constant
background model. The free parameters are the normalization of the PSF
and the background level.  The fits were unacceptable: $\chi^2$ = 20.9
and 30.4 for 11 degrees of freedom, in the 0.5--2 keV and 2--7 keV
bands, respectively. In order to parameterize the spatial extent, we
fitted the radial profiles with a constant background plus a
two-dimensional Gaussian convolved through the PSF. The free
parameters in this fit are the dispersion and normalization of the
Gaussian and the background level. This model gave a significantly
better fit with $\Delta \chi^2$=14 and 21, respectively, for one
additional parameter. These are significant at more than 99\%
confidence. The best-fit values of $\sigma$ are summarized in Table 4.
The Gaussian model provides a reasonably good representation of the
observed profile. These results indicate that the X-ray images are
extended on kpc scales in both the 0.5--2 keV and 2--7 keV bands. The
difference in spatial extent between the two energy bands is not
significant. The best-fit profile is shown in Figure 2.

\subsection{NGC 4192}

NGC 4192 and a few serendipitous sources were detected in the field.
The contour maps of the SIS images in the 0.5--7 keV band are shown in
Figure 1{\it b}. The positions of the detected sources are summarized
in Table 3. An archival {\rosat} PSPC image shows two sources, which
combine to yield an elongated morphology, centered on the galaxy. The
position angle of the elongation is $\sim72$ deg and these sources are
aligned roughly along the direction of the minor axis of the
galaxy. In the {\asca} image, this elongation is not clearly seen
because of limited photon statistics and spatial resolution.

Since NGC 4192 is very dim and its exposure time was shorter than
those of other objects, we fitted the one-dimensional projection of
the SIS images, as in the case of NGC 404, to measure the X-ray
fluxes; the projection was made with a width of 3\farcm2 along the
nucleus and the serendipitous source (source 3 in Table 3). We fitted
the resulting profile with a constant background plus two PSFs
centered at the position of the galaxy nucleus and source 3.  The free
parameters in the fit are the normalization of each PSF and the
background level.  The fit of the projected profiles in the 0.5--2 keV
and 2--10 keV bands yielded X-ray fluxes of $5.7\times10^{-14}$ ergs
s$^{-1}$ cm$^{-2}$ and $1.1 \times10^{-13}$ ergs s$^{-1}$ cm$^{-2}$,
respectively, for an assumed power-law spectrum with $\Gamma$ = 1.7
(see \S\ 4). At an adopted distance of 16.8 Mpc (Tully 1988), these
fluxes correspond to the luminosities of $L$(0.5--2 keV) =
$1.9\times10^{39}$ ergs s$^{-1}$ and $L$(2--10 keV) =
$3.8\times10^{39}$ ergs s$^{-1}$.  The significance of the detection
is 9.6$\sigma$ and 6.7$\sigma$ in the 0.5--2 keV and 2--10 keV bands,
respectively.

\subsection{NGC 4457}

NGC 4457 was detected in both the SIS and GIS images. Contour maps of
GIS images in the 0.5--2 keV and 2--7 keV bands are shown in Figure
1{\it e} and 1{\it f}. Serendipitous sources are also detected in the
GIS field of view, and their positions are summarized in Table 3. The
brightest one, located at ($\alpha$,~$\delta$)$_{\rm J2000}$ =
($12^h\,29^m\, 47^s$, 3\deg\ 35\amin\ 32\asec), is plausibly
identified with the Virgo cluster galaxy VCC 1208
[($\alpha$,~$\delta$)$_{\rm J2000}$ = ($12^h\,29^m\,39.2^s$, 3\deg\
36\amin\ 43\asec)].

We fitted the radial profiles of the SIS images in the 0.5--2 keV and
2--7 keV bands using the same procedure as in NGC 4111. The PSF fits
gave $\chi^2$ values of 26.9 and 15.3, respectively, for 11 degrees of
freedom. The results of the Gaussian fits are shown in Table 4. The
image in the hard band is consistent with being pointlike, but the
upper limit on $\sigma$ is large ($\sigma<1.8$ arcmin). Although the
lower boundary of the Gaussian $\sigma$ for the soft-band image fit is
greater than zero, this result cannot rule out the possibility that
the soft X-ray source is pointlike, since it is possible that Gaussian
fits to a point source results in nonzero $\sigma$ (\ltsima 0\farcm2)
(Ptak 1997). The best-fit profile is shown in Figure 2.

\subsection{NGC 4569}

NGC 4569 was detected in both the SIS and GIS images, and a few
serendipitous sources were found in the GIS field of view. The GIS
images in the 0.5--2 keV and 2--7 keV bands are shown in Figures 1{\it
g} and 1{\it h}. The position of the brightest source
[($\alpha$,~$\delta$)$_{\rm J2000}$ = ($12^h\,37^m\,34^s$, 13\deg\
18\amin\ 47\asec)] coincides closely with that of the QSO Q1235+1335
[($\alpha$,~$\delta$)$_{\rm J2000}$ = ($12^h\,37^m\,33.6^s$, 13\deg\
19\amin\ 6\farcs6); $z$=0.15].  Diffuse emission from the hot gas in
the Virgo cluster is also seen in the soft-band image. NGC 4569 is
separated from M87 by 2.1 degrees, and the cluster emission at this
angular distance has been detected in a {\it ROSAT} PSPC image by
B\"ohringer et al. (1994).

We compared the radial profiles of the SIS images in the 0.5--2 keV
and 2--7 keV bands with those of the PSF and found that the SIS images
are clearly extended in both energy bands. The PSF fits yielded
$\chi^2$ = 37.2 and 30.7 for 11 degrees of freedom, respectively. The
best-fit $\sigma$ for the Gaussian model is shown in Table 4, and the
profiles are shown in Figure 2. The $\chi^2$ improved significantly in
this model for one additional parameter ($\Delta \chi^2$ = 22 and 24
for the 0.5--2 keV and 2--7 keV images, respectively). The profile in
the 2--7 keV band is well fitted by a Gaussian with $\sigma$ =
1\farcm6.  On the other hand, the residuals of the fit in the 0.5--2
keV band suggest the presence of a compact source at the center in
addition to the emission extended over arcminute scales. These two
components can be identified with the unresolved emission seen in a
{\rosat} HRI image (Colbert \& Mushotzky 1999) and the extended
emission detected in a {\rosat} PSPC image (Junkes \& Hensler 1996).

\placefigure{fig1}
\placetable{tbl-3}

\placefigure{fig2}
\placetable{tbl-4}

\section{X-ray spectra}

We fitted the X-ray spectra of NGC 4111, 4117, 4457, and 4569. The
spectra were extracted using a circular region centered on the nucleus
with a radius of 3\amin--4\amin which is consistent with the sizes of
objects. There were no confusing sources within the extraction
radius. Background spectra were extracted from a source-free region in
the same field. The spectra from the two SIS detectors (SIS0 and SIS1)
were combined, as were those from the two GIS detectors (GIS2 and
GIS3).  Then we fitted the SIS and GIS spectra simultaneously. The
Galactic hydrogen column densities used in the spectral fits are
derived from the \ion{H}{1} observations by Dickey \& Lockman
(1990). Since NGC 4192 is too faint for a detailed spectral analysis,
we estimated its spectral shape using the hardness ratio, defined to
be the photon flux ratio between the 2--10 keV and 0.5--2 keV bands.

\subsection{NGC 4111, NGC 4457, and NGC 4569}

The X-ray spectra of NGC 4111, NGC 4457 and NGC 4569 cannot be fitted
with a single-component model. We tried a power-law model and a
thermal bremsstrahlung model. The absorption column density of the
matter along the line of sight was treated as a free parameter.  We
obtained unacceptable fits (Table 5). Since the value of the obtained
reduced $\chi^2$ is large ($>2$), errors are not shown for spectral
parameters in Table 5. A bump is seen around 0.8--0.9 keV in all the
spectra, which can be identified with the Fe~L line complex.  This
feature suggests the presence of a thermal plasma with a temperature
of $\sim 0.7$ keV. A Raymond-Smith (hereafter RS; Raymond \& Smith
1977) thermal plasma model also failed to give an adequate fit, and
significant positive residuals were seen above $\sim1.5$ keV. This
indicates the presence of a hard component in addition to the soft
thermal emission.  Accordingly, we fitted the spectra with a
two-component model which consists of a soft thermal component and a
hard component. We used the RS plasma to represent the soft thermal
component and a power-law or thermal bremsstrahlung contribution as
the hard component. We assumed a Galactic value for the absorption
column density of the RS component.  The abundances were fixed at 0.1
of the solar value since it would otherwise not be well constrained
from the data; adopting other values (0.3 and 0.5 solar) gave similar
results within the errors. The absorption column density for the hard
component is treated as free parameter. Table 5 summarizes the results
of the fitting.  The two-component model reproduces well the observed
spectra, and both a power law and a thermal bremsstrahlung model yield
similarly good fits. The observed spectra and the best-fit RS plus
power-law models are shown in Figure 3, and Table 7 lists the derived
X-ray luminosities. The RS plus thermal bremsstrahlung model gives
very similar luminosities.

It is worth noting that the temperatures of the thermal bremsstrahlung
component in NGC 4111 and NGC 4457 are not well constrained, and only
lower limits of the temperature are obtained.  Additionally, if the
abundance of the RS component were allowed to vary, only NGC 4111
gives a constrained value (0.006--1.0 solar). We could set only a
lower limit of the abundance for NGC 4457 ($>$0.001 solar) and NGC
4569 ($>$0.05 solar).

The X-ray fluxes obtained from SIS and GIS are consistent with each
other, within the errors, for NGC 4111 and NGC 4457.  In the case of
NGC 4569, the SIS and GIS fluxes differ at the level of 20\%--50\%;
the SIS gives $\sim$ 20\% smaller flux in the soft band below 2 keV,
while its flux in hard band above 2 keV is $\sim$ 50\% larger.  This
discrepancy is possibly due to the diffuse emission in the NGC 4569
field and to the different background extract regions. In the spectra
of NGC 4569, negative residuals are seen around 1.2 keV. This may be
also due to imperfect subtraction of surrounding diffuse emission,
since the temperature of the Virgo intracluster gas in this region is
$\sim$ 2.5 keV, and both the Fe~L line complex and the He-like Mg line
are expected around an energy of 1.1--1.3 keV (B\"ohringer et
al. 1994; Matsumoto 1998). Because the SIS and GIS each has a
different spatial resolution and field of view, we cannot perfectly
match the source-free region used for background subtraction.
Therefore, we regard this discrepancy (up to 50\%) as a systematic
error inherent in the derive X-ray fluxes and luminosities of NGC 4569. 
The typical errors for the fluxes and luminosities of NGC 4111 and NGC
4457 are $\sim$25\% (this value does not include the calibration
uncertainty of $\sim$10\%).

\subsection{NGC 4192}

Since NGC 4192 is very faint, we estimated its spectral slope using a
hardness ratio, $f$(2--10 keV)/$f$(0.5--2 keV). The photon flux in
each energy band was calculated from the fits of the projected image
described in \S\ 3.3. We obtained a hardness ratio of $0.48\pm0.09$
(1$\sigma$ errors). If we assume a spectral shape of a power law
absorbed by the Galactic hydrogen column density in the direction of
this galaxy ({\NH} = $2.7\times10^{20}$ {\pcm}), this hardness ratio
corresponds to $\Gamma = 1.70^{+0.19}_{-0.16}$ (1$\sigma$ errors).

\subsection{NGC 4117}

For completeness, we mention the results of the spectral analysis for
NGC 4117, a low-luminosity Seyfert 2 galaxy serendipitously observed
in the GIS field of NGC 4111. It is of great interest to compare the
spectral properties of LINER 2s with low-luminosity Seyfert 2s.  The
GIS spectrum of NGC 4117 is shown in Figure 4. It is clear that the
soft X-rays are significantly absorbed by a large column density.  We
fitted the spectrum with an absorbed power-law model. The best-fit
parameters are: $\Gamma = 0.92^{+1.16}_{-0.81}$ and {\NH} =
$2.8^{+1.6}_{-1.0}\times10^{23}$ {\pcm}. Since small positive
residuals are seen below 2 keV, we also tried to add a power-law
component with little absorption. We assumed that the photon indices
of both power laws are the same (equivalent to a partially covered
power-law model) and that the absorption column for the less absorbed
power-law component is equal to the Galactic value ({\NH} =
$1.4\times10^{20}$ {\pcm}). In this model, $\chi^2$ improved by only
$\Delta \chi^2 = -2.5$, and the resulting best-fit model parameters,
as summarized in Table 6, are $\Gamma = 1.11^{+0.97}_{-1.01}$ and
{\NH} = $3.0^{+0.9}_{-1.1}\times10^{23}$ {\pcm}. The X-ray luminosities
for this model are shown in Table 7, where we adopt a distance of 17
Mpc (Tully 1988). The intrinsic X-ray luminosity of $1.3\times10^{41}$
{\eps} is one of the lowest values ever observed for Seyfert 2
galaxies in the hard X-ray band.

We added a narrow Gaussian to the above models to constrain the Fe~K
fluorescent line. No significant improvement of $\chi^2$ was
obtained. The upper limits of the equivalent width are 150 and 125 eV
for the single power-law model and the partially covered power-law
model, respectively.

The observed X-ray spectrum obscured by large column density ({\NH}
$>10^{23}$ {\pcm}) is quite similar to more luminous Seyfert 2
galaxies. The obtained best-fit photon index is flat (0.9--1.1),
although the error is large. It is flatter than the canonical value in
Seyfert 1 galaxies (e.g. Nandra et al. 1997) and such an apparently
flat spectral slope is often observed in Seyfert 2 galaxies (Awaki et
al. 1991a; Smith \& Done 1996; Turner et al. 1997a). Seyfert 2
galaxies usually show fluorescent Fe~K emission line. Although the
upper limit on the equivalent width is slightly smaller than that
expected from cold matter of {\NH} = $3\times10^{23}$ {\pcm} along the
line of sight (150--200 eV; Awaki et al. 1991a; Leahy \& Creighton
1993; Ghisellini, Haardt, \& Matt 1994), it is consistent with Seyfert
2 galaxies within the scatter in the plot of equivalent width versus
absorption column density (Fig 1 in Turner et al. 1997b). Therefore, we
found no clear difference between luminous Seyfert 2 galaxies and the
low luminosity Seyfert 2 NGC 4117.

\placefigure{fig3}
\placetable{tbl-5}

\placefigure{fig4}
\placetable{tbl-6}

\placetable{tbl-7}

\section{Discussion}

\subsection{Hard Component}

We have detected hard X-ray emission from all the objects except for
NGC 404.  The hard X-ray (2--7 keV) images of NGC 4111 and NGC 4569
are clearly extended on scales of several kpc, an indication that a
nonstellar, active nucleus is not the primary source of the hard X-ray
emission. This is also consistent with the lack of time variability.

Other lines of evidence suggest that at least some of these galaxies
have experienced recent star formation.  The UV spectra of NGC 404 and
NGC 4569, for example, show unambiguous stellar absorption lines
arising from young, massive stars (Maoz et al. 1998).
Starburst galaxies are also a source of hard X-rays, and their X-ray
spectra can be modeled as thermal bremsstrahlung emission with a
temperature of several keV (e.g., Moran \& Lehnert 1997; Ptak et
al. 1997; Persic et al. 1998). However, the morphologies of the hard
X-ray emission in starburst galaxies tend to be either pointlike or
only slightly extended (Tsuru et al. 1997; Cappi et al. 1999),
significantly more compact than observed in our sample.  It appears,
therefore, that hard X-ray emission associated with starburst activity
does not significantly contribute to the emission observed in the
objects in our sample, although this conclusion remains at the moment
tentative because only a small number of starburst galaxies have been
studied in the hard X-rays.  Note that the extended hard X-ray
emission in the starburst galaxy M83 is interpreted as due to a
collection of X-ray binaries in its bulge (Okada, Mitsuda, \& Dotani
1997).

In normal spiral galaxies, X-ray emission comes mainly from discrete
sources such as low-mass X-ray binaries (Fabbiano 1989; Makishima et
al. 1989). The X-ray size of NGC 4111 and NGC 4569 in the hard band is
similar to the optical size. The upper limit on the X-ray size of NGC 4457
is also consistent with its optical size. The extended hard-band
images of the above objects are consistent with a discrete source
origin. Their X-ray spectra, approximated by a thermal bremsstrahlung
model with a temperature of several keV, are as expected from a
collection of low-mass X-ray binaries (Makishima et al. 1989). The
X-ray luminosities of normal spiral galaxies are roughly proportional
to the their optical ($B$ band) light (Fabbiano 1989).  Table 8 gives
the {\LX}/{\LB} values for the galaxies in our sample; taking into
consideration the scatter in the {\LX}--{\LB} relation, these values
agree with those seen in normal galaxies.  We conclude that the origin
of the extended hard X-ray emission in our sample probably arises from
a collection of discrete X-ray sources in the host galaxy, and we find
no clear evidence for the presence of an AGN.

\placetable{tbl-8}

Note, however, that our data do not rule out the presence of an
AGN. If present, the nonstellar X-ray luminosity in the 2--10 keV band
must be significantly smaller than few $\times$ $10^{39}$ {\eps}.  In
the case that an AGN core is heavily obscured by a large column
density of {\NH} $>\, 10^{24}$ {\pcm}, then we expect only to see
X-ray emission scattered by a warm and/or cold reflector.  If the
scattering fraction is $\sim$3\% (Turner et al. 1997b; Awaki, Ueno, \&
Taniguchi 1999), an upper limit on the intrinsic luminosity is
estimated to be $\sim10^{41}$ {\eps}. Recent {\it BeppoSAX}
observations have shown that several X-ray weak Seyfert 2s are highly
obscured by Compton-thick matter (Maiolino et al. 1998). Such a
situation could also occur in LINER 2s. If a heavily obscured AGN is
present, a strong Fe~K emission line at 6.4 keV is expected with an
equivalent width larger than 1 keV (e.g., Terashima et
al. 1998b). Unfortunately, this hypothesis cannot be tested with our
data because the limited photon statistics do not permit us to set
stringent upper limits on the equivalent width of the Fe~K line
(typical upper limits of the equivalent widths are $\sim$ 2 keV). We
will be able to address this issue with future high-energy
observations on missions that have larger effective areas, such as
{\it XMM} and {\it ASTRO-E}, and that have much finer spatial
resolution, such as {\it Chandra}.

Recent work on massive black holes in (nearly) normal galaxies have
shown that the mass of the central black hole is about 1/100 to 1/1000 of
the mass of the bulge (e.g., Magorrian et al. 1998). Using this black
hole mass - bulge mass correlation and our upper limits on X-ray
luminosities of AGN, we can estimate upper limits of the Eddington
ratio of the accretion onto massive black holes in the galaxies in our
sample. We calculated the black hole mass using the relations $M_{\rm
BH}$ = 0.005$M_{\rm bulge}$ and $M_{\rm bulge}$ = $5\times10^9
M_{\odot}(L_{\rm bulge}/10^9L_{\odot})^{1.2}$ (Richstone et al. 1998
and references therein). 

We used the upper limits on the X-ray luminosities in the 2--10 keV
band, of $1.5\times10^{39}$ ergs s$^{-1}$ for NGC 404 and
$1\times10^{41}$ ergs s$^{-1}$ for the others, and assumed a
bolometric correction of a factor of 10 and a scattering fraction of
$\sim3 \%$. The bulge luminosities are calculated by using the data in
Table 11 in Ho et al. 1997a. We obtained upper limits on the Eddington
ratios in the range of $(1-5)\times10^{-5}$. Thus we found that mass
accretion is taking place with very low accretion rate or at very low
efficiency if super massive black holes are present in these galaxies,
as is in most galaxies.

\subsection{Soft Component}

In NGC 4111, NGC 4457, and NGC 4569 we detected soft X-ray emission
that can be represented by a Raymond-Smith thermal plasma with
$kT\approx0.65$ keV.  Extended hot gas with a temperature $kT<1$ keV
is generally observed in starburst galaxies, and it is interpreted as
due to gas shock heated by the collective action of supernovae (e.g.,
Dahlem, Weaver, \& Heckman 1998). The X-ray luminosity of the hot gas
component is roughly proportional to the far-infrared (FIR)
luminosity: $\log$~{\LX}/{\LFIR} $\approx$ --4 (Heckman, Armus, \&
Milley 1990; David, Forman, \& Jones 1992). We calculated the
{\LX}/{\LFIR} ratios for the galaxies in our sample in order to test
the starburst origin of the soft thermal component. We use the X-ray
luminosities of the RS component corrected for absorption in the
0.5--4 keV band, and {\LFIR} is calculated from the flux
$1.26\times10^{-14}(2.58S_{60}+S_{100})$ W m$^{-2}$, where $S_{60}$
and $S_{100}$ are the flux densities at 60 $\mu$m and 100 $\mu$m,
respectively, in units of janskys (Table 1; see Ho et al. 1997a for
details).  We used observed total luminosities in the 0.5--2 keV band
for NGC 404 and NGC 4192, since we cannot separate the thermal
component from the total emission. Relations between soft X-ray
luminosities and far infrared luminosities are plotted in Figure 5,
where Soft X-ray luminosities are only for a Raymond-Smith plasma
component except for NGC 404 and NGC 4192. We compiled X-ray
luminosities of the soft thermal component in X-ray bright starburst
galaxies in {\it ASCA} archives (for detailed results, see Okada et
al. 1996; Ptak et al. 1997; Dahlem, Weaver, \& Heckman 1998; Ptak et
al. 1999; Della Ceca et al. 1999; Zenzas et al. 1999; Heckman et
al. 1999; Moran, Lehnert, \& Helfand 1999). We fitted their {\it ASCA}
spectra with a model consisting of a RS plasma plus a thermal
bremsstrahlung and measured intrinsic (absorption corrected) X-ray
luminosities of the RS component. These points are also shown in
Figure 5. The $\log$~{\LX}/{\LFIR} values (Table 8 and Figure 5)
distribute around a value of $\sim$--3.5 -- --4, which is consistent, within
the scatter, with what is seen in starburst galaxies. With {\it ASCA}
spatial resolution we can only say that the X-ray size is roughly the
same as the optical size, consistent with that indicated by higher
resolution {\it ROSAT} data for starburst galaxies.

The radial profile of the soft X-ray image of NGC 4569 indicates the
presence of a compact nuclear component in addition to the extended
component ($\sigma>1$\amin).  A comparison between the radial profiles
in the soft and hard bands suggests that the compact component has a
soft spectrum which could be identified with an unresolved source seen
in the {\rosat} HRI image of Colbert \& Mushotzky (1999). A compact
but resolved nuclear source is detected in a UV image of NGC 4569
taken with {\HST} (Barth et al. 1998), and absorption features in its
UV spectrum indicate that the UV source is a cluster of hot stars
(Maoz et al. 1998).  The X-ray spectra of O-type stars can be modeled
by a thermal plasma with a temperature of $kT\sim$ 1 keV (e.g.,
Corcoran et al. 1994; Kitamoto \& Mukai 1996). Since individual O
stars have X-ray luminosities of $10^{32}-10^{33}$ {\eps} (e.g.,
Rosner, Golub, \& Vaiana 1985), about $10^7$ O stars would be required
to explain the X-ray source at the nucleus detected in the {\rosat}
HRI image and in the {\asca} soft-band image. This number, however, is
too large compared to the number of O stars needed to explain the
observed H$\alpha$ luminosity and the strength of the UV continuum
($<$1000; Maoz 1999). Moreover, we can rule out the possibility of
such a giant cluster of O stars from dynamical constraints.  A cluster
of $10^7$ O5 stars, each $\sim$40 \solmass, would amount to a total
mass of $4\times 10^8$ \solmass.  Even under the unreasonable
assumption that stars of lower masses are absent, this mass strongly
violates the dynamical mass limit of the nucleus, which has been
estimated to be $< 10^6 - 10^7$ \solmass\ by Keel (1996).  Therefore,
we conclude that the hot star contribution to the observed soft X-ray
luminosity is minor. This argument also applies to all the objects in
our sample, except for NGC 404, for which only an upper limit of X-ray
luminosity is available.

\placefigure{fig5}

Finally, we note that scattered light from a hidden AGN is unlikely to
be the source of the soft X-ray emission. If AGN emission is blocked
by a large column density along the line of sight, only emission
scattered by cold and/or warm material would be observed. If we fit
the X-ray spectra below 2 keV with a simple power-law model, the
photon indices become steeper than 3 for NGC 4111, NGC 4457, and NGC
4569. This spectral slope is significantly steeper than normally seen
in Seyfert 2s, where scattering by warm material is thought to
prevail. Cold reflection produces an X-ray spectrum that is flatter
than the intrinsic spectrum.  Additionally, the extended morphologies
of the soft X-ray emission makes it unlikely that the photoionized
medium can be maintained at a highly ionized state. Thus the soft
component probably does not originate from scattered AGN emission.
Instead, the most likely origin for the soft thermal component is
supernovae-heated hot gas.

\subsection{Ionization Photon Budget}

We have found no clear evidence for the presence of an AGN in the
galaxies in our sample. In order to compare the ionization source in
LINER 2s and transition objects with those of LINER 1s and
low-luminosity Seyfert galaxies, we calculated {\LX}(2--10 keV)/{\LHa}
ratios (Table 8), where we used the luminosities of the narrow
component of the H$\alpha$ emission.  The {\LX}/{\LHa} values of the
galaxies in our sample are systematically lower, by more than one
order of magnitude, compared to LINER 1s and low-luminosity Seyferts
(Terashima 1999; Terashima et al. 1999), objects where LLAGNs are
almost certainly present.  The mean $\log$~{\LX}/{\LHa} is 1.6 for
LLAGNs, while the $\log$~{\LX}/{\LHa} values for our sample are
smaller than 0.61.

We estimate the number of ionizing photons needed to account for the
observed H$\alpha$ luminosities, assuming a spectral energy
distribution of $f_{\nu} \propto \nu^{-1}$, which is the typical
spectral shape between the UV and X-rays observed in LLAGNs (Ho
1999b), Case B recombination (Osterbrock 1989), and a covering factor
of unity for the ionized gas. A value of $\log$~{\LX}/{\LHa} = 1.4 is
sufficient to explain the H$\alpha$ luminosity by photoionization by
an AGN.  The observed $\log$~{\LX}/{\LHa} values, on the other hand,
are significantly lower than this. If the AGN is not significantly
obscured, its luminosity in the 2--10 keV band is estimated to be less
than a few $\times$ $10^{39}$ {\eps}. In this case, photons from an
AGN account for only a very small fraction ($\sim$5\%) of the observed
H$\alpha$ luminosity. If H$\alpha$ is due to ionization by an AGN, one
would then have to postulate that the AGN is heavily obscured with a
column density greater than $\sim 10^{24}$ {\pcm} and that only
scattered radiation is observable.  Alternatively, an ionizing source
other than an AGN is required.

As discussed in the next subsection, galaxies in our sample have lower
[OI]$\lambda6300$/H$\alpha$ values than LINERs that are most likely to
be AGN.  If the low [\ion{O}{1}]/H$\alpha$ ratio is due to dilution by
\ion{H}{2} regions, which produce strong Balmer lines and very weak
[OI], the difference in {\LX}/{\LHa} between the two classes is
reduced. Since the median of the [OI]/H$\alpha$ ratio for the LINER 2s
is about a factor of 3 smaller than that for the LINER 1s with LLAGNs,
the H$\alpha$ emission tracing the AGN would be 1/3 of the total
measured value. Even in this case, however, the observed hard X-ray
luminosities are not enough to drive the H$\alpha$ luminosities.

\subsection{Optical Emission Line Ratios and Ionizing Source}
                            
Our sample includes three LINERs (NGC 404, NGC 4111, and NGC 4457) and
two transition objects (NGC 4192 and NGC 4569). By the definition of
Ho et al. (1993, 1997a), transition objects have a smaller [O~I]
$\lambda$6300/H$\alpha$ ratio than LINERs (Table 8; data from Ho et
al. 1997a). This class of emission-line nuclei has been postulated to
be composite systems where a LINER nucleus is spatially contaminated
by circumnuclear star-forming regions (Ho et al. 1993; Ho 1996). On
the other hand, photoionization by hot stars in environments with
ionization parameters characteristically lower than in ``normal''
giant extragalactic \hii\ regions also generates the spectral
properties of transition objects (Filippenko, \& Terlevich 1992;
Shields 1992). The presence of hot stars is seen directly in the UV
spectrum of NGC 4569 (Maoz et al. 1998); these stars can provide the
power to explain the observed emission-line luminosities if very
massive stars are still present.  The low X-ray output of these
systems, as found in this study, lends further support for this
conclusion.  Thus, at least some transition objects are likely to be
powered by hot stars. The nucleus of NGC 4192, however, is not
detected in the UV band, possibly because of the large extinction due
to the high inclination of the galaxy (83\deg; Barth et al. 1998).
The X-ray properties and optical emission-line ratios of NGC 4192 are
similar to those of NGC 4569, and it, too, might be primarily
powered by hot stars.

We compared the [O~I]/H$\alpha$ ratios of the LINERs in our sample
with those of LINERs from which AGN-like X-ray emission has been
detected. The [O~I]/H$\alpha$ ratios for NGC 404, NGC 4111, and NGC
4457 (Table 8) are lower than in LINERs that are strong LLAGN
candidates, and they are located at the lowest end of the distribution
of [O~I]/H$\alpha$ in LINERs (Fig. 7 in Ho et al. 1997a).  For
comparison, the [O~I]/H$\alpha$ values for the few LINERs where
compact hard X-ray emission has been detected are 0.71 (NGC 1052),
0.53 (NGC 3998), 1.22 (NGC 4203), 0.48 (NGC 4579), 0.18 (NGC 4594),
and 0.24 (NGC 4736).\footnote{The X-ray results for these objects are
published in Weaver et al. (1999), Guainazzi, \& Antonelli (1999),
Awaki et al. (1991b), Iyomoto et al. (1998), Terashima et al. (1998a),
Nicholson et al. (1998), and Roberts et al. (1999).}  It is intriguing
that the {\it HST} UV spectrum of NGC 404 also shows strong evidence
for the presence of energetically significant hot stars (Maoz et
al. 1998); no UV spectral information is available for NGC 4111 and
NGC 4457.  It is conceivable that the subset of LINERs with
exceptionally weak [O~I] emission owe their primary excitation
mechanism to stellar photoionization.  Obviously more observations are
necessary to settle this issue.  A statistical study using a large
sample of objects will be presented elsewhere.

\section{Summary}

We presented \asca\ results for a small sample of LINERs (NGC 404, NGC
4111, and NGC 4457) and transition objects (NGC 4192 and NGC
4569). X-ray emission was detected in all objects except NGC 404. The
X-ray luminosities in the 2--10 keV band range from $4\times10^{39}$
to $1\times10^{40}$ {\eps}. The images of NGC 4111 and NGC 4569 are
extended on scales of several kpc in both the soft ($<$2 keV) and hard
($>$2 keV) energy bands. The X-ray spectra of NGC 4111, NGC 4457, and
NGC 4569 are well represented by a two-component model consisting of a
soft thermal plasma of $kT\sim0.65$ keV plus a hard component (power
law or thermal bremsstrahlung).

The soft X-ray emission probably originates from hot gas produced via
recent star formation activity because both the temperature of the gas
and the {\LX}/{\LFIR} ratios are typical of starburst galaxies. The
extended morphology of the hard X-ray emission indicates that it
mainly comes from discrete sources in the host galaxies, and that the
AGN contribution is small, if any. The {\LX}(2--10 keV)/{\LHa} values
for the galaxies in our sample are more than one order of magnitude
smaller than in LINERs with bona fide LLAGNs (those with a detectable
broad H$\alpha$ emission line and compact hard X-ray emission), and
the X-ray luminosities are insufficient for driving the optical
emission-line luminosities. These facts imply that, if an AGN is
present, it would have to be heavily obscured with a column density
much greater than {\NH} $\approx\, 10^{23}$ {\pcm}.  We suggest that
the optical emission lines in the galaxies in our sample are mainly
powered by photoionization by hot, young stars rather than by an AGN.
This hypothesis is supported by the detection of stellar features due
to massive stars in the UV spectra of NGC 404 and NGC 4569, by the
systematically lower [O~I]/H$\alpha$ ratios in these objects compared
to LINERs with bona fide LLAGNs, and by the low observed X-ray
luminosities reported in this work.

We also analyzed the X-ray properties of NGC 4117, a low-luminosity
Seyfert 2 galaxy serendipitously observed in the field of NGC 4111,
and found its properties to be consistent with other Seyfert 2
galaxies with moderate absorbing columns.

\acknowledgments

The authors are grateful to all the {\asca} team members. We also
thank an anonymous referee for useful comments. YT thanks JSPS for
support. LCH acknowledges partial financial support from NASA grants
GO-06837.01-95A, GO-07357.02-96A, and AR-07527.02-96A, which have been
awarded by the Space Telescope Science Institute (operated by AURA,
Inc., under NASA contract NAS5-26555).  We made use of the NASA/IPAC
Extragalactic Database (NED) which is operated by the Jet Propulsion
Laboratory, California Institute of Technology, under contract with
NASA.


\newpage

\begin{center}
{\bf Figure Captions}
\end{center}

\figcaption[]{
({\it a}) SIS image of NGC 404 from 0.5--2 keV, 
({\it b}) SIS image of NGC 4192 from  0.5--7 keV, 
({\it c}) GIS image of NGC 4111 from  0.7--2 keV, 
({\it d}) GIS image of NGC 4111 from  2--10 keV, 
({\it e}) GIS image of NGC 4457 from  0.7--2 keV, 
({\it f}) GIS image of NGC 4457 from  2--7 keV, 
({\it g}) GIS image of NGC 4569 from  0.7--2 keV, 
({\it h}) GIS image of NGC 4569 from  2--7 keV.
The crosses indicate the optical position of the nucleus taken from
the NED. The contours are linearly spaced. The lowest contour levels are 
(a)40\%, (b) 40\%, (c) 25\%, (d) 15\%, (e) 40\%, (f) 45\%, 
(g) 40\%, and (h) 45\% of the peak. Background is not subtracted in 
these images. Sky coordinates are J2000.
\label{fig1}
}

\figcaption{ Radial surface brightness profiles fitted with a two- dimension
Gaussian convolved through the PSF + constant model.  {\it Solid} lines are 
the best-fit model and {\it dotted} lines are the background level. 
{\it Dot-dashed} lines are the PSF normalized at the innermost data point. 
({\it a}) NGC 4111, 0.5--2 keV, ({\it b}) NGC 4111, 2--7 keV, ({\it c}) 
NGC 4457, 0.5--2 keV, ({\it d}) NGC 4457, 2--7 keV, ({\it e}) NGC 4569, 
0.5--2 keV, ({\it f}) NGC 4569, 2--7 keV.
\label{fig2}
}

\figcaption{
({\it a}) SIS spectrum of NGC 4111, 
({\it b}) GIS spectrum of NGC 4111,
({\it c}) SIS spectrum of NGC 4457, 
({\it d}) GIS spectrum of NGC 4457, 
({\it e}) SIS spectrum of NGC 4569, 
({\it f}) GIS spectrum of NGC 4569.
Crosses are the observed data and histograms are the best-fit
Raymond-Smith + absorbed power-law model. The Raymond-Smith 
component is plotted with a {\it dot-dashed} line, and the power-law 
component is plotted with a {\it dashed} line.
\label{fig3}
}

\figcaption{
GIS spectrum of NGC 4117. Crosses are the observed data and histograms 
are the best-fit absorbed power-law model.
\label{fig4}
}

\figcaption[]{Correlation between soft X-ray luminosity and fir
infrared luminosity for type 2 LINERs (filled circles) and starburst
galaxies (crosses). The soft X-ray luminosities are the Raymond-Smith
component of a two component spectral fit to the {\asca} data except for
NGC 404 and NGC 4192. Dashed lines correspond to $\log L_{\rm
X}$/$L_{\rm FIR}$ = --3 (upper) and -- 4 (lower). Starburst galaxy
sample consists of NGC 5236 (M83), NGC 253, NGC 3310, NGC 3034(M82),
NGC 2146, NGC 3256, and NGC 3690 (from left to right in the
plot). \label{fig5} }

\newpage


\footnotesize

\begin{table*}
	\caption{Observed galaxies\label{tbl-1}}
\begin{center}
\begin{tabular}{ccccccccccc}
\tableline \tableline
Name	& Distance$^a$	& Classification & $\log F$(H$\alpha$) & $\log L$(H$\alpha$)	& $\log L_{FIR}$	& $\log L_{B}$\\
	& [Mpc]		&		& [{\eps} {\pcm}] 	& [{\eps}]		& [{\eps}]	& [$L_{\odot}$]\\
\tableline
NGC 404	&  2.4		& L2		& --13.21	& 37.63	& 40.94	& 8.58\\
NGC 4111 & 17.0		& L2		& --13.14	& 39.40	& ---	& 10.01\\
NGC 4192 & 16.8		& T2		& --13.56	& 38.97	& 43.25	& 10.64\\
NGC 4457 & 17.4		& L2		& --12.99	& 39.57	& 42.98	& 10.05\\
NGC 4569 & 16.8		& T2		& --12.36$^b$	& 40.174$^b$ & 43.34& 10.73\\
\tableline
\end{tabular}
\end{center}
\tablecomments{a: Tully (1988); b: Keel (1983);
Other data are taken from Ho et al. 1997a.}
\end{table*}

\begin{table*}
	\caption{Observation log\label{tbl-2}}
\begin{center}
\begin{tabular}{lllcccc}
\tableline \tableline
Name	& Date	& SIS mode	& \multicolumn{2}{c}{SIS} & \multicolumn{2}{c}{GIS}\\
	&	&		& count rate	& exposure & count rate	& exposure\\
	&	&		& [counts s$^{-1}$] & [ksec] &  [counts s$^{-1}$] & [ksec]\\
\tableline
NGC 404	&1997 Jul 21 	&1CCD Faint & ---	& 12.3	& ---	& 13.4\\
	&1998 Feb 6 	&1CCD Faint & ---	& 27.2	& ---	& 29.7\\
NGC 4111&1997 Dec 7 	&1CCD Faint LD0.48keV$^a$ & 0.008	& 14.4	& 0.004	& 15.6\\
	&1997 Dec 15 	&1CCD Faint LD0.48keV$^a$ & 0.008	& 19.0	& 0.003	& 21.2\\
NGC 4192&1997 Dec 17	&2CCD Faint/Bright & ---& 3.9	& ---	& 4.3\\
	&1997 Dec 23	&2CCD Faint/Bright & ---& 13.3	& ---	& 14.3\\
NGC 4457&1998 Jun 14--15&1CCD Faint & 0.01	& 42.1	& 0.005	& 44.0\\
NGC 4569&1997 Jun 24--25&1CCD Faint & 0.03	& 21.9	& 0.006	& 21.0\\
	&1997 Jul 6-7	&1CCD Faint & 0.03	& 19.0	& 0.007	& 20.4\\
\tableline
\end{tabular}
\end{center}
\tablecomments{$^a$: Level discriminator is enabled.}
\end{table*}

\begin{table}[hbt]
        \caption{Positions of serendipitous sources\label{tbl-3}}
\begin{center}
\begin{tabular}{lll}
\hline \hline
source No.& position (J2000)            & ID\\
\hline
NGC 404\\
1*	& 1h 9m 28s, 35d 38m 52s	& \\
NGC 4111\\
0*	& 12h 7m  6s, 43d 3m 16s	& NGC 4111 12h07m03.2s  +43d03m55.3s\\
1	& 12h 7m 47s, 43d 6m 53s	& NGC 4117 12h07m46.1s  +43d07m35.0s\\
2	& 12h 8m 16s, 43d 3m 12s	& \\
NGC 4192\\
0*	& 12h 13m 49s, 14d 53m 47s	& NGC 4192 12h13m48.3s +14d54m01s\\
1	& 12h 13m 5s, 14d 54m 57s	& \\
2	& 12h 13m 9s, 14d 51m 8s        & LBQS 1210+1507  12h13m08.0s +14d51m06s\\
3	& 12h 14m 12s, 14d 47m 31s	& \\
NGC 4457\\
0*	& 12h 28m 56s, 3d 34m 15s	& NGC 4457 12h28m59.2s  +03d34m16.1s\\
1	& 12h 28m 17s, 3d 38m 43s	& \\
2	& 12h 28m 28s, 3d 25m 43s	& \\
3	& 12h 29m 2s, 3d 38m 14s	& \\
4	& 12h 29m 47s, 3d 35m 32s	& VCC 1208 12h29m39.2s  +03d36m43s \\
NGC 4569\\
0*      & 12h36m50s, +13d10m01s       & NGC 4569 (12h36m49.8s, +13d09m46s)\\
1       & 12h36m30s, +13d07m59s       & \\
2       & 12h36m34s, +12d58m54s       & \\
3       & 12h37m34s, +13d18m47s       & QSO Q1235+1335 (12h37m33.6s, +13d19m6.6s)\\
\hline
\end{tabular}
\end{center}
\tablecomments{*: position determined using SIS images. Typical error radius
is 1 arcmin.}
\end{table}

\begin{table}[hbt]
        \caption{Gaussian fit to radial profiles\label{tbl-4}}
\begin{center}
\begin{tabular}{cccccc}
\hline \hline
Name	& Scale		& \multicolumn{2}{c}{0.5--2 keV}	& \multicolumn{2}{c}{2--7 keV}	\\
	&		& $\sigma$	& $\chi^2$/dof	& $\sigma$	& $\chi^2$/dof\\
	& [kpc/arcmin]	& [arcmin]	&		& [arcmin]	&		\\
\hline
NGC 4111 & 4.95		& 0.50(0.27-0.89) & 6.6		& 1.2(0.37-3.3)	& 9.2\\
NGC 4457 & 5.06		& 0.43(0.15-0.65) & 18.2	& 0.11(0-1.8)	& 15.3\\
NGC 4569 & 4.89		& 1.4(1.0-1.8)	& 15.7		& 1.6(1.0-3.5)	& 6.54 \\
\hline
\end{tabular}
\end{center}
\tablecomments{The quoted errors in parentheses are at the 90\% confidence
level for one interesting parameter. The degree of freedom in the fit is 10.}
\end{table}

\begin{table*}
	\caption{Results of spectral fits\label{tbl-5}}
\begin{center}
\begin{tabular}{clcccccc}
\tableline \tableline
Name	& Model	& \NH $_1^a$ & $kT^b$	& abundance$^c$	& \NH $_2^d$	& photon index$^e$	& $\chi^2$/dof \\
	&	& [$10^{22}${\pcm}] & [keV]	& [solar] & [$10^{22}${\pcm}] &	or $kT$ [keV]	& \\
\tableline
NGC 4111& Brems	& 	&		&		& 0		& 0.65			& 74.3/24\\
	& PL	&	&		&		& 0		& 3.1			& 48.7/24\\
     & RS+Brems	& 0.014(f)& $0.65^{+0.12}_{-0.14}$& 0.1(f)	& $0.49^{+2.3}_{-0.49}$ & $>5.0$	& 14.9/22\\
	& RS+PL	& 0.014(f)& $0.65^{+0.12}_{-0.14}$& 0.1(f)	& $0^{+3.0}$	& $0.9^{+1.1}_{-0.6}$	& 14.4/22\\
NGC 4457 &Brems	& 	&		&		& 0		& 3.1			& 70.7/27\\ 
	& PL	& 	&		&		& 0		& 2.3			& 50.1/27\\
     & RS+Brems	& 0.018(f)& $0.68^{+0.12}_{-0.16}$ & 0.1(f)	& $0.84^{+2.2}_{-0.84}$ & $>2.8$	& 17.9/25\\
	& RS+PL & 0.018(f)& $0.68^{+0.12}_{-0.16}$ & 0.1(f)	& $1.1^{+2.7}_{-1.1}$	& $1.7^{+1.2}_{-0.8}$ & 18.0/25\\
NGC 4569 &Brems	& 	&	 	&		& 0		& 3.9			& 114.0/35\\
	& PL	&	&		&		& 0		& 2.0			& 98.0/35\\
     & RS+Brems & 0.029(f)& $0.67^{+0.09}_{-0.14}$ & 0.1(f)	& $1.2^{+0.8}_{-0.9}$	& $5.4^{+25}_{-2.5}$	& 56.2/32\\
	& RS+PL	& 0.029(f)& $0.67^{+0.09}_{-0.10}$ & 0.1(f)	& $1.7^{+1.0}_{-1.1}$	& $2.2\pm0.7$	& 55.0/32\\
\tableline
\end{tabular}
\end{center}
\tablecomments{The quoted errors in parentheses are at the 90\% confidence
level for one interesting parameter. (f) denotes a frozen parameter. The 
model components are as follows: PL = power-law with absorption; Brems = 
thermal bremsstrahlung with absorption; RS = Raymond-Smith thermal plasma
with the Galactic absorption.
$^a$: Galactic foreground absorption column density, 
$^b$: temperature of the Raymond-Smith component,
$^c$: abundance of the Raymond-Smith component,
$^d$: absorption column density of the hard component
$^e$: photon index of the power-law component or temperature of the thermal
Bremsstrahlung component.
}
\end{table*}

\begin{table*}
\caption{Spectral fitting results of NGC 4117\label{tbl-6}}
\begin{center}
\begin{tabular}{ccccc}
\tableline \tableline
Model	& \NH	& Photon index	& covering fraction	& $\chi^2$/dof\\
		& [$10^{22}${\pcm}] &	&			& \\
\tableline
Power-law	& $28^{+16}_{-10}$  & $0.92^{+1.16}_{-0.81}$ 	& & 21.9/17\\
Partially covered power-law	& $0.014^a, 30^{+9 b}_{-11}$& $1.11^{+0.97}_{-1.01}$ & $0.984^{+0.016}_{-0.038}$ & 19.4/15\\
\tableline
\end{tabular}
\end{center}
\tablecomments{
(f) denotes a frozen parameter.
$^a$: absorption column density of the uncovered (less absorbed) component dominating the lower energy range,
$^b$: absorption column density of the covered component.
}
\end{table*}

\begin{table*}
	\caption{X-ray luminosities\label{tbl-7}}
\begin{center}
\begin{tabular}{lcccccc}
\tableline \tableline
Name	& Total (2--10 keV) & (0.5--2 keV)	& \multicolumn{2}{c}{Power-law (2--10 keV)}	& \multicolumn{2}{c}{Raymond-Smith (0.5--4 keV)}\\
	& observed	& observed	& observed	& intrinsic	& observed	& intrinsic \\
\tableline
NGC 404	& $<0.046$ 	& $<0.0076$	& ---	& ---	& ---	& ---\\
NGC 4111 & 9.0		& 7.1	& 8.7	& 8.7	& 6.0	& 6.3\\
NGC 4192 & 3.8		& 1.9	& ---	& ---	& ---	& ---\\
NGC 4457 & 9.1		& 6.4	& 8.8	& 9.6	& 5.6	& 5.9\\
NGC 4569 & 10		& 8.9	& 9.8	& 12	& 8.0	& 8.9\\
NGC 4117 & 46		& 0.44	& 46	& 130	& ---	& ---\\
\tableline
\end{tabular}
\end{center}
\tablecomments{unit: $10^{39}$ {\eps}. A RS+PL model is assumed for 
NGC 4111, 4457, and 4569. The luminosities for NGC 404 and NGC 4192 
are obtained from the projected image fitting. A partially covered 
power-law model is assumed for NGC 4117. Upper limits for NGC 404 are 
3$\sigma$.}
\end{table*}

\begin{table*}
\caption{Luminosity ratios\label{tbl-8}}
\begin{center}
\begin{tabular}{lcccc}
\tableline \tableline
& $\log L_{\rm HX}/L_{{\rm H}\alpha}$
& $\log L_{\rm HX}/L_B$ 
& $\log L_{\rm SX}/L_{\rm FIR}$ 
& [O~I]/H$\alpha$ \\
\tableline 
NGC 404		&$<0.03$& $<-4.52$& $<-4.06$ & 0.17\\
NGC 4111	& 0.54	& --3.66& ---	& 0.19\\
NGC 4192	& 0.61	& --4.65& --3.67& 0.14\\
NGC 4457	& 0.41	& --3.66& --3.21& 0.19\\
NGC 4569	&--0.11 & --4.26& --3.39& 0.062\\
\tableline
\end{tabular}
\end{center}
\tablecomments{$L_{\rm HX}$: Intrinsic luminosity of the
power-law component in 2--10 keV; $L_{\rm SX}$: Intrinsic
luminosity of the Raymond-Smith component in 0.5--4 keV for NGC 4457
and NGC 4569, and observed luminosity in 0.5--2 keV for NGC 404 and
NGC 4192}
\end{table*}

\clearpage

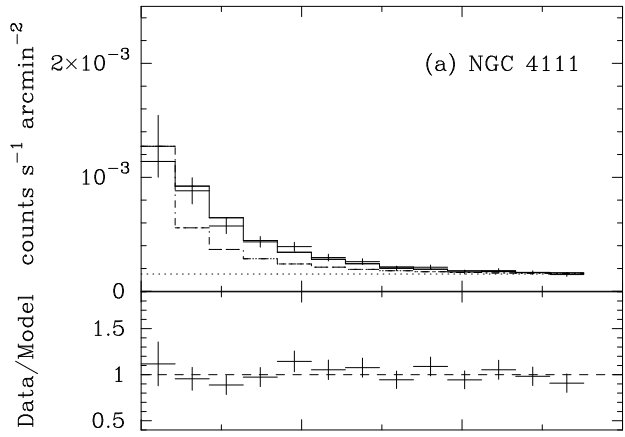
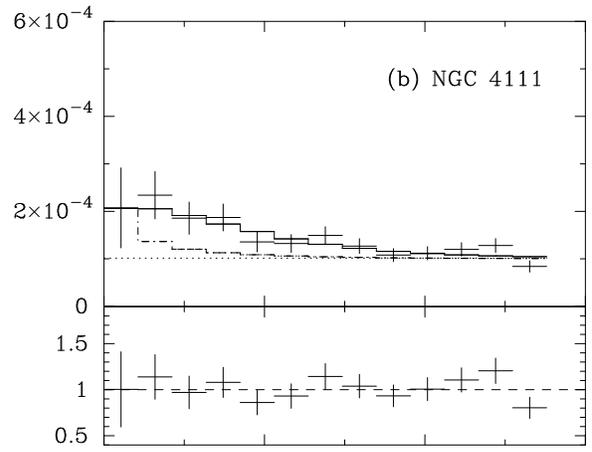
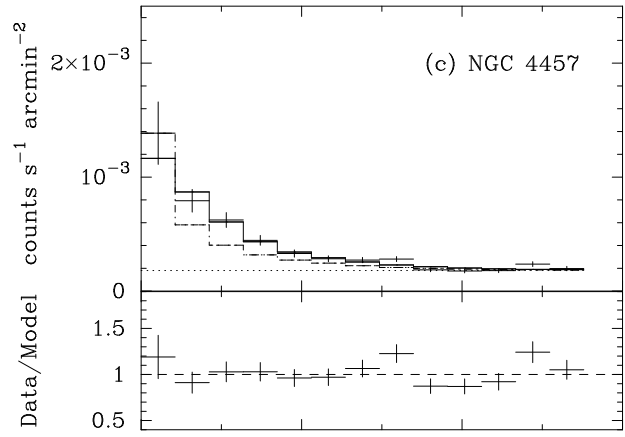
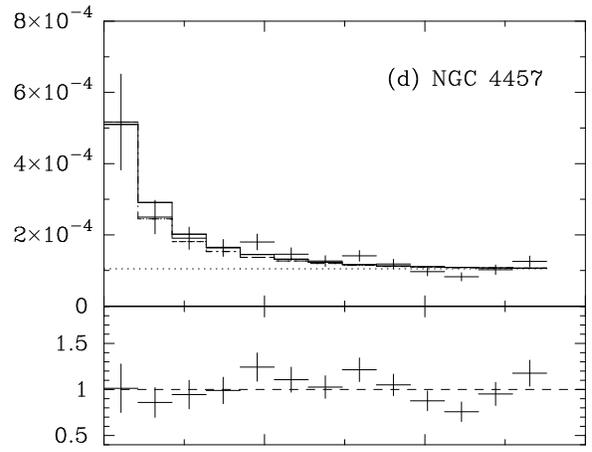
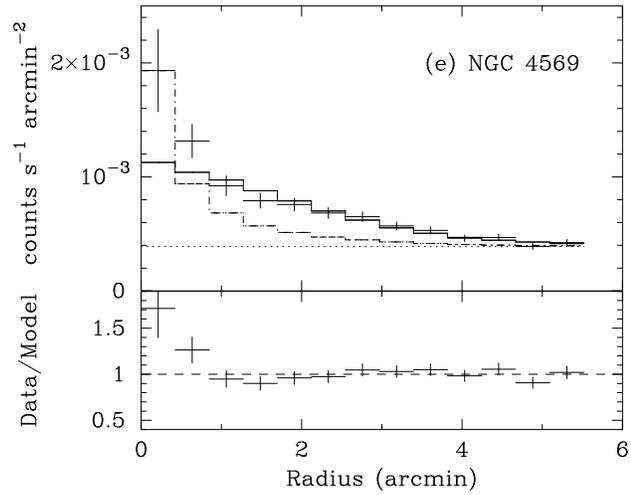
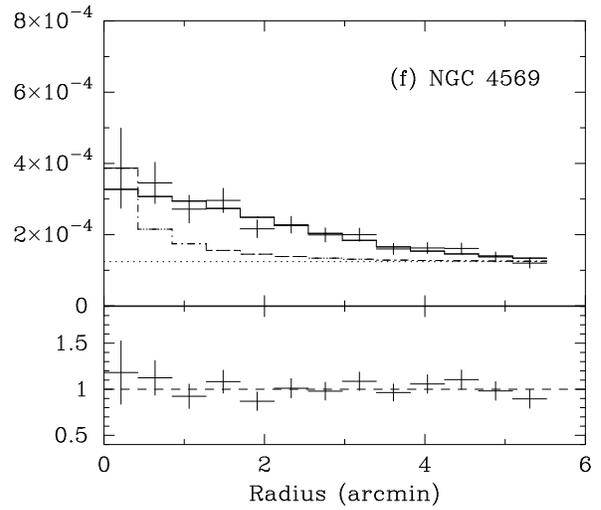
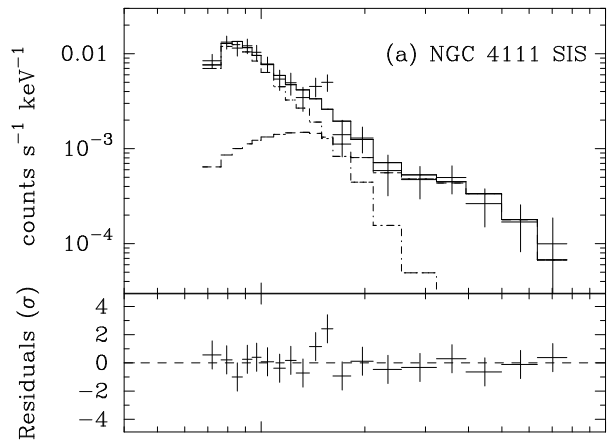
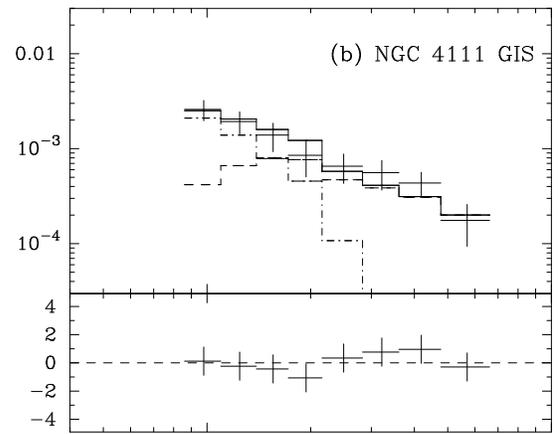
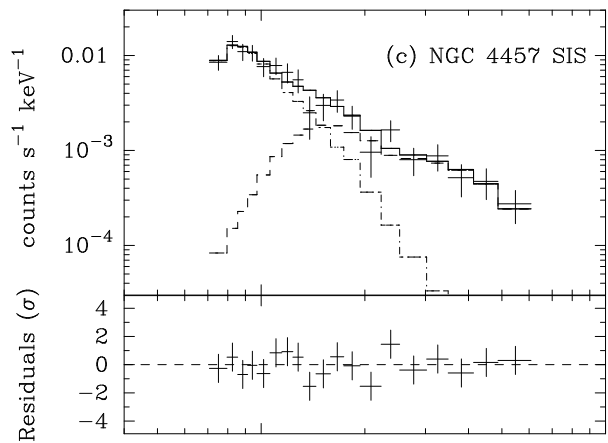
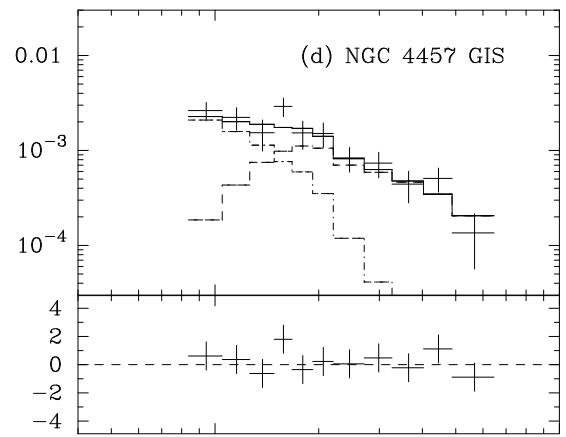
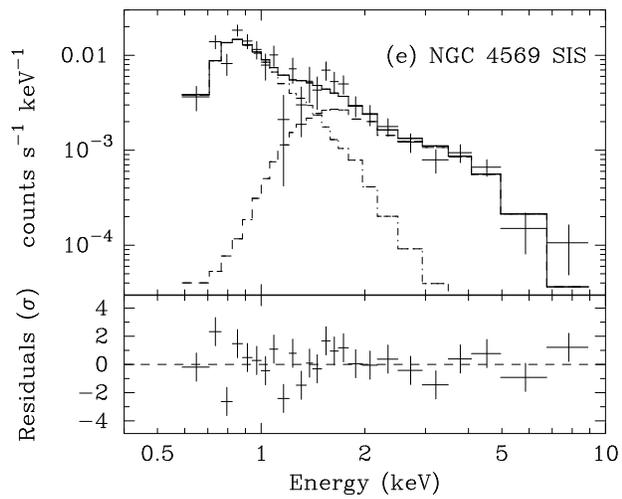
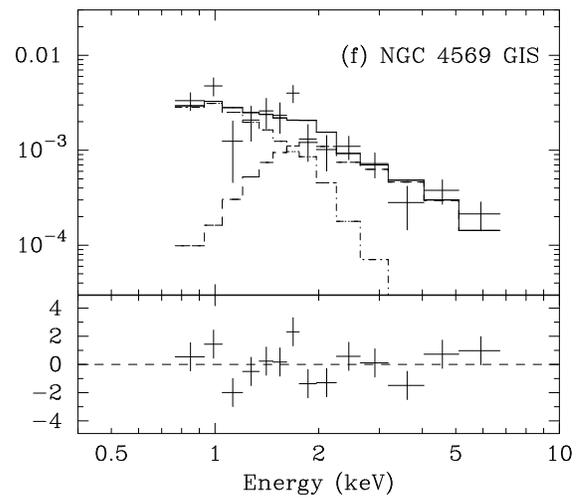
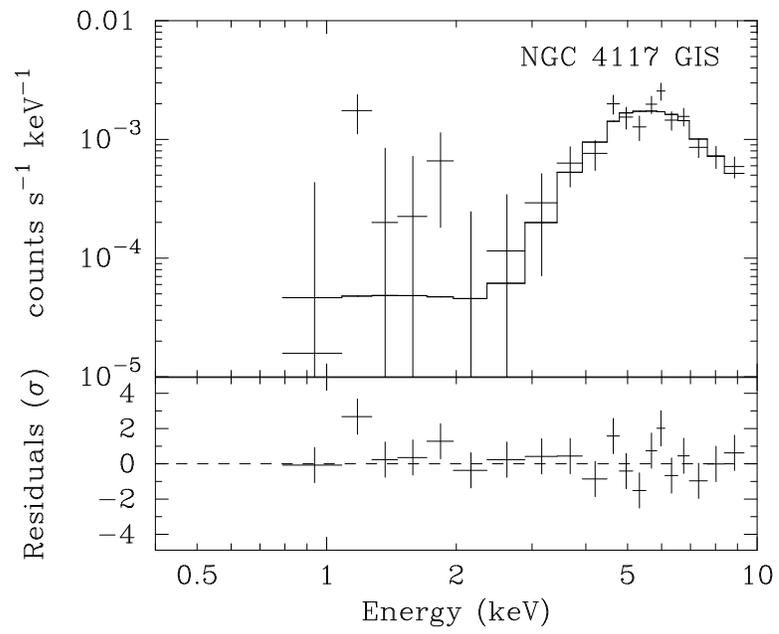
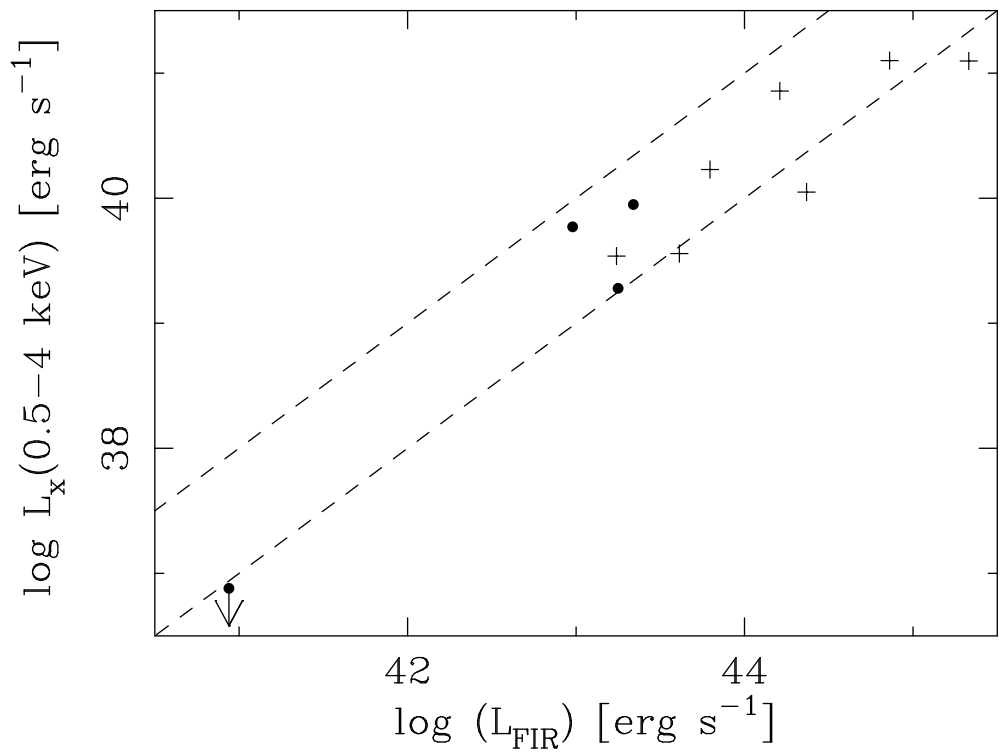
\end{document}